\documentclass[showpacs,pra,twocolumn]{revtex4}
\usepackage{amsfonts}
\usepackage{amsmath}
\usepackage{amssymb}
\usepackage{graphicx}

\begin{document}

\title{Phase Separation in two-Species Atomic Bose-Einstein Condensate with
Interspecies Feshbach Resonance}
\author{Lu Zhou$^{1}$, Jing Qian$^{1}$, Han Pu$^{2,\ast \ast }$, Weiping
Zhang$^{1,\dagger }$, and Hong Y. Ling$^{3,\ast }$}
\affiliation{$^{1}$State Key Laboratory of Precision Spectroscopy, Department of Physics,
East China Normal University, Shanghai 200062, China}
\affiliation{$^{2}$Department of Physics and Astronomy, and Rice Quantum Institute, Rice
University, Houston, TX 77251-1892, USA }
\affiliation{$^{3}$Department of Physics and Astronomy, Rowan University, Glassboro, New
Jersey 08028-1700, USA}

\begin{abstract}
We consider a mixture of two-species atomic Bose-Einstein codensates coupled
to a bound molecular state at zero temperature via interspecies Feshbach
resonance. The interspecies Feshbach coupling precludes the possibility of
doubly mixed phases while enables not only the pure molecular superfluid but
also the pure atomic superfluids to exist as distinct phases. We show that
this system is able to support a rich set of phase separations, including
that between two distinct mixed atom-molecule phases. We pay particular
attention to the effects of the Feschbach coupling and the particle
collisions on the miscibility of this multi-component condensate system.
\end{abstract}

\maketitle

\section{Introduction\ }

The experimental realization of multi-component condensate systems in the
late-90s \cite{myatt97,hall98,stenger98} has renewed the interest in the
subject of miscibility of two distinguishable condensates in a binary
mixture \cite{ho96,ao98,bohn97,pu98,timmermans98}. Whether two condensates
coexist in the same spatial volume (miscibility) or\ repel each other into
separate spatial regions (immiscibility or phase separation) or even
collapse when they are brought together is largely determined by the
relative strength between inter- and intra-species s-wave scattering
lengths. The ability to change the two-body interactions by tuning the
magnetic field across the Feshbach resonance in atomic systems has enabled
the observation of phase separations, that are otherwise impossible with
traditional condensed matter environments, to be accessed experimentally.
The recent experimental demonstration of rich phase structure including the
phase separation in an unbalanced fermionic superfluidity system \cite%
{ketterle06,hulet06} serves as a great testament to the remarkable
controllability that the technique of Feshbach resonance can bring to the
atomic system.

In this paper, we consider a model where bosonic atoms of two different
species are coupled to a bound heteronuclear molecular state via the
interspecies Feshbach resonance (we will simply refer to it as the
heteronuclear model in this paper). Chiefly due to their large permanent
electric dipole moments, ultracold heteronuclear molecules are thought to
hold great promise in applications such as quantum computing and simulation
\cite{demille02,zoller06}, and precision measurement of fundamental symmetry
\cite{kl95,hudson02,demille08} and constants \cite%
{flambaum07,zelevinsky08,demille081}. As such, the study of cold
heteronuclear molecules has evolved into one of the most hotly pursued
researches in the field of atomic, molecular and optical physics \cite%
{hutson06}. The recent experimental efforts both in achieving the tunable
interaction between $^{41}$K and $^{87}$Rb \cite{inguscio08} and in
demonstrating the ability of the Feshbach resonance to control the
miscibility of a $^{85}$Rb-$^{87}$Rb dual-species BEC \cite{wieman08} have
further stimulated our interest in the proposed model.

In an earlier paper \cite{zhou07}, we have concentrated on the coherent
creation of heteronuclear molecular condensates (similar studies can also be
found in reference \cite{duncan07}\textbf{)}. In this paper, we will focus
on constructing zero-temperature phase diagrams, paying particular
attentions to phase separation and collapse. \ In models with a broad
Feshbach resonance, the molecular component can be adiabatically eliminated;
this approximation leads to an effective binary condensate system, whose
properties were investigated extensively in the past\ \cite%
{ho96,ao98,bohn97,pu98,timmermans98}. In general, we must treat the
molecular component as an independent degree of freedom, regarding our model
as a genuine three-species condensate system. However, our system is
fundamentally different from the independent gas model \cite{ueda06,yabu07},
where no interspecies conversions can take place. The conversion between
atoms and molecules can lead to phenomena unconventional to the independent
gas model, as has been pointed out by Radzihovsky \textit{et al}. \cite%
{radzihovsky04} and Romans \textit{et al}. \cite{romans04}. The surprise,
that the homonuclear model admits the molecular superfluid (MSF) but not the
pure atomic superfluid (ASF), was traced to the \textit{intraspecies}
Feshbach\ process which takes \textit{two }atoms from the same species to
form \textit{one} molecule. This is no longer the case for the \textit{%
interspecies} Feshbach process which requires only \textit{one} atom from
each species to produce \textit{one} molecule. But, it creates a new
surprise - the doubly mixed phases with two nonvanishing order parameters
which are ubiquitous in the independent gas model are completely absent from
our system. For the same reason, not only the pure molecular phase but also
the pure atomic superfluids can exist in the heteronuclear model. This along
with the possibility of simultaneously tuning the Feshbach energy and the
atomic population imbalance (an extra control \textquotedblleft
knob\textquotedblright\ unique to the heteronuclear model) makes the number
of experimentally accessible phase separations (summarized in Sec. VI) far
greater in the heteronuclear model than in the homonuclear model.

This paper is organized as follows. In Sec. II, we describe our model and
the Hamiltonian by defining all the system parameters. In Sec. III, we
derive the equilibrium conditions by minimizing a total grand canonical
Hamiltonian under the premise that the system is in an arbitrary state of
phase separation. A general classification of various phases supported by
our system is summarized in Sec. IV. The same section also presents the
nonlinear coupled equation governing the mixed atom-molecule superfluid
phase, and generalize the stability conditions to our three-component
system. In Sec. V, we provide numerical examples that showcase the ability
of our system to support a rich set of phase separation possibilities while
at the same time elucidate the effects of the Feshbach coupling and the
particle collisions on the miscibility property of our system. Finally, a
summary is given in Sec. VI.

\section{Model and Hamiltonian}

Our model consists of two distinct atomic species of state $\left\vert
1\right\rangle $ and $\left\vert 2\right\rangle $, and a molecular species
of state $\left\vert 3\right\rangle $. The state $\left\vert 3\right\rangle $
differs in energy from the atomic states by an amount $\delta $. In our
model, particles can interact with each other through collisions. At low
temperature, we use $\chi _{ij}$ to denote the collisional strength of the
s-wave scattering involving either same species $\left( i=j\right) $ or
different species $\left( i\neq j\right) $. In addition, atoms of state $%
\left\vert 1\right\rangle $ can combine with atoms of state $\left\vert
2\right\rangle $ to coherently create molecules of state $\left\vert
3\right\rangle $ via Feshbach resonance characterized with atom-molecule
coupling strength $g$ and Feshbach detuning $\delta $, tunable by changing
the magnetic field. \textbf{\ }Let $\hat{\Psi}_{i}\left( \mathbf{r}\right) $%
\ be the field operator for the $i$-th component, and $m_{i}$\ be the mass
of species $i$\ with $m_{3}=m_{1}+m_{2}$. We describe\ our system with the
following Hamiltonian%
\begin{eqnarray}
\hat{H} &=&\int d^{3}\mathbf{r}\left\{ \sum_{i=1,2,3}\hat{\Psi}_{i}^{\dagger
}\left( \mathbf{r}\right) \left( -\frac{\hbar ^{2}\nabla ^{2}}{2m_{i}}%
\right) \hat{\Psi}_{i}\left( \mathbf{r}\right) \right.   \notag \\
&&+\delta \hat{\Psi}_{3}^{\dagger }\left( \mathbf{r}\right) \hat{\Psi}%
_{3}\left( \mathbf{r}\right) +\frac{g}{2}\left[ \hat{\Psi}_{3}^{\dagger
}\left( \mathbf{r}\right) \hat{\Psi}_{1}\left( \mathbf{r}\right) \hat{\Psi}%
_{2}\left( \mathbf{r}\right) +h.c\right]   \notag \\
&&\left. +\sum_{i,j}\frac{\chi _{ij}}{2}\hat{\Psi}_{i}^{\dagger }\left(
\mathbf{r}\right) \hat{\Psi}_{j}^{\dagger }\left( \mathbf{r}\right) \hat{\Psi%
}_{j}\left( \mathbf{r}\right) \hat{\Psi}_{i}\left( \mathbf{r}\right)
\right\} ,  \label{H}
\end{eqnarray}%
which can be easily generalized from the Hamiltonian for the homonuclear
model \cite{radzihovsky04,romans04}.

In this paper, we limit our study to the homogeneous condensate system at
zero-temperature where each species has already condensed into the zero
momentum mode. As such, we can describe such a system in the spirit of the
mean-field approximation in which $\hat{\Psi}_{i}\left( \mathbf{r}\right) $
is replaced with the uniform $c$-number $\psi _{i}\exp \left( i\varphi
_{i}\right) $, where $\psi _{i}>0$ and $\varphi _{i}$ are two real numbers,
representing, respectively, the modulus and the phase of the corresponding
complex order parameter. \ It can be easily shown that the mean-field energy
density $\mathcal{E=}\langle \hat{H}\rangle /V$ (with $V$ being the total
volume) is a function of the "relative phase" $\varphi =\varphi _{1}+\varphi
_{2}-\varphi _{3}$ and is minimized when $\varphi =\pi $ (assuming, without
loss of generality, that $g>0$). With these considerations, we derive from
Eq. (\ref{H}) the following mean-field (low branch) energy density%
\begin{equation}
\mathcal{E}=\delta \psi _{3}^{2}-g\psi _{1}\psi _{2}\psi _{3}+\sum_{ij}\frac{%
\chi _{ij}}{2}\psi _{i}^{2}\psi _{j}^{2},  \label{Ehomo}
\end{equation}%
which will be used in the next section as the starting point for studying
phase and phase separations.

\section{General Conditions for Phase Separation}

To begin with, we imagine that our system at the ground state is phase
separated into a number of arbitrary phases (yet to be determined), each of
which occupying a volume of $V_{i}$. For the purpose of easy illustration,
without loss of generality, we consider the case that the system separates
into two different phases. The total energy of the system, when the
interface energies are ignored \cite{ao98,timmermans98}, then becomes%
\begin{equation}
E_{total}=V_{1}\mathcal{E}^{\left( 1\right) }+V_{2}\mathcal{E}^{\left(
2\right) },  \label{E total}
\end{equation}%
where $\mathcal{E}^{\left( n\right) }$ is the mean-field energy density of
the $n$th phase and is equivalent to $\mathcal{E}$ in Eq. (\ref{Ehomo}) when
$\psi _{i}$ is replaced with $\psi _{i}^{\left( n\right) },$ the order
parameter of the $i$th component in the $n$th phase. \ The ground state
energy without phase separation is then a special case where, for example, $%
V_{2}=0$ and $V_{1}=V$.

To determine the ground state, we construct a generalized (mean-field) grand
canonical Hamiltonian%
\begin{equation}
K_{total}=E_{total}-\mu N-hN_{d}-mV  \label{K total}
\end{equation}%
where $\mu $, $h$, and $m$ are three Lagrangian multipliers associated,
respectively, with the conservation of the total particle number
\begin{equation}
\sum_{n=1}^{2}\left[ \left( \psi _{1}^{\left( n\right) }\right) ^{2}+\left(
\psi _{2}^{\left( n\right) }\right) ^{2}+2\left( \psi _{3}^{\left( n\right)
}\right) ^{2}\right] V_{n}=N.  \label{N conservation}
\end{equation}%
the total atom number difference
\begin{equation}
\sum_{n=1}^{2}\left[ \left( \psi _{1}^{\left( n\right) }\right) ^{2}-\left(
\psi _{2}^{\left( n\right) }\right) ^{2}\right] V_{n}=N_{d}  \label{Nd}
\end{equation}%
and finally the total volume%
\begin{equation}
V_{1}+V_{2}=V.  \label{V conservation}
\end{equation}

To proceed further, we first define the free energy density
\begin{equation}
\mathcal{K}=-\sum_{i}\mu _{i}\psi _{i}^{2}-g\psi _{1}\psi _{2}\psi
_{3}+\sum_{ij}\frac{\chi _{ij}}{2}\psi _{i}^{2}\psi _{j}^{2}.  \label{Ka}
\end{equation}%
When the use of Eqs. (\ref{E total}), (\ref{N conservation}), and (\ref{V
conservation}) is made, we find that%
\begin{equation}
K_{total}=V_{1}\mathcal{K}^{\left( 1\right) }+V_{2}\mathcal{K}^{\left(
2\right) }-m\left( V_{1}+V_{2}\right) ,  \label{K total 1}
\end{equation}%
where $\mathcal{K}^{\left( n\right) }$ is equivalent to $\mathcal{K}$ in
Eq.~(\ref{Ka}) with $\psi _{i}$ and $\mu _{i}$ substituted with $\psi
_{i}^{\left( n\right) }$ and $\mu _{i}^{\left( n\right) }$, respectively. In
Eq.~(\ref{Ka}),
\begin{equation*}
\mu _{i}^{\left( 1\right) }=\mu _{i}^{\left( 2\right) }=\mu _{i},
\end{equation*}%
with $\mu _{1}=\mu +h$, $\mu _{2}=\mu -h$, and $\mu _{3}=2\mu -\delta $.
Thus, as expected, the chemical potentials in equilibrium are balanced as a
consequence of the particle number conservation laws [Eqs. (\ref{N
conservation}) and (\ref{Nd})]. \ Minimization of $K_{total}$ with respect
to $V_{1}$ and $V_{2}$ results in
\begin{equation}
\mathcal{K}^{\left( 1\right) }=\mathcal{K}^{\left( 2\right) }.
\end{equation}

Thus, we conclude that (1) for such a phase separation to be possible, phase
1 and phase 2 must coexist for a given set of $\mu _{1}$, $\mu _{2}$ and $%
\mu _{3}$ or must overlap in the chemical potential space, and (2) phase
separation takes place along the first-order phase transition line where the
energy density of phase 1 equals that of phase 2 \cite{radzihovsky07}.
Evidently, the same conclusion holds for phase separation involving an
arbitrary number of phases.

\section{Possible Phases and Stability Criteria}

\subsection{Possible Phases}

\label{Sec:possiblePhase}

In order to determine the possible phases, we minimize Eq.~(\ref{K total 1})
with respect to $\psi _{i}^{\left( n\right) }$. This procedure results in
the following set of saddle-point equations,

\begin{eqnarray}
\mu _{1}\psi _{1} &=&-\frac{g}{2}\psi _{2}\psi _{3}+\psi _{1}\sum_{i}\chi
_{1i}n_{i},  \notag \\
\mu _{2}\psi _{2} &=&-\frac{g}{2}\psi _{1}\psi _{3}+\psi _{2}\sum_{i}\chi
_{2i}n_{i},  \notag \\
\mu _{3}\psi _{3} &=&-\frac{g}{2}\psi _{1}\psi _{2}+\psi _{3}\sum_{i}\chi
_{3i}n_{i}.  \label{psi123}
\end{eqnarray}%
where we have replaced $\psi _{i}^{\left( n\right) }$ with $\psi _{i}$ and
used $n_{i}=\psi _{i}^{2}$ to denote the density of the $i$th species. The
possible phases, which correspond to different solutions of Eqs.~(\ref%
{psi123}), can then be broadly divided into five groups: \ \

(i) Vacuum with $\psi _{1}=\psi _{2}=\psi _{3}=0;$

(ii) Atomic superfluid of species 1 (ASF1) with $\psi _{1}=\sqrt{\mu
_{1}/\chi _{11}}$, $\psi _{2}=\psi _{3}=0;$

(iii) Atomic superfluid of species 2 (ASF2) with $\psi _{2}=\sqrt{\mu
_{2}/\chi _{22}}$, $\psi _{1}=\psi _{3}=0;$

(iv) Molecular superfluid (MSF) with $\psi _{3}=\sqrt{\mu _{3}/\chi _{33}}$,
$\psi _{1}=\psi _{2}=0;$

(v) Mixed atom-molecule superfluid (AMSF) with $\psi _{1}\neq 0$, $\psi
_{2}\neq 0$ and $\psi _{3}\neq 0$.

We first note that our model allows the existence of both MSF and ASF, in
contrast to the homonuclear model, which supports MSF but not ASF. In our
model, the energy density due to the Feshbach coupling is $g\psi _{1}\psi
_{2}\psi _{3}$, which is symmetric with respect to the exchange of any pair
of states and hence does not favor the formation of one pure phase to
another. \ \ In the homonuclear model, on the other hand, the energy density
due to the Feshbach coupling is in the asymmetric form of $g\psi _{3}\psi
_{1}^{2}$. With such a form, condensation of atoms will inevitably lead to
the formation of the molecular condensate while condensation of molecules
can take place in the absence of the atomic population.

We emphasize that this classification holds only in the presence of the
Feshbach coupling $\left( g\neq 0\right) $. In a typical three interacting
gas system where $g=0$, the system supports, besides all the phases listed
above, doubly mixed phases, for example, ASF1-MSF in which $\psi _{1}$ $\neq
0$, $\psi _{3}\neq 0$, and $\psi _{2}=0$. The absence of the doubly mixed
phases can again be attributed to the unique form of the Feshbach coupling, $%
g\psi _{1}\psi _{2}\psi _{3}$, in the heteronuclear model. This term
dictates that the existence of condensate population in any pair of states
will lead to the condensation in the third component, which precludes the
possibility of forming the doubly mixed phases where the population in the
third component is zero.

We also comment that for a given set of chemical potentials, while ASF1,
ASF2, and MSF are unique, AMSF, due to the highly nonlinear nature of Eq. (%
\ref{psi123}), may itself be divided into different types depending on the
density distribution among three species. Such an example will be given in
Sec. \ref{sec:twoAMSFs}. \

Finally,\ this classification, although obtained under the condition of zero
temperature, can provide a useful road map for understanding the phase
diagrams at finite temperature. At finite temperature, state (i) becomes the
so-called normal state where the gas mixture does not possess any order, and
other states become mixed in the sense that superfluidities and thermal
densities are both present. \ Lowering temperature reduces the entropy and
restores the order to the gas system, thereby giving rise to a hierarchy of
critical temperatures associated with phase transitions that break
symmetries at various levels. The subject of finite temperature phase
diagrams will be left as a future study.

\subsection{Stability Criteria}

From Eq. (\ref{psi123}) we have determined the possible phases, which
correspond to the critical points (extrema or saddle points) of the free
energy density $\mathcal{K}$. We now turn our attention to studying their
stability properties by identifying the local minimum of the function $%
\mathcal{K}$. For this purpose, we first introduce the Hessian matrix
\begin{equation}
Hessian=\left(
\begin{array}{ccc}
f_{11} & f_{12} & f_{13} \\
f_{12} & f_{22} & f_{23} \\
f_{13} & f_{23} & f_{33}%
\end{array}%
\right) ,  \label{hessian}
\end{equation}%
where the matrix element is defined as $f_{ij}=\partial _{\psi _{i}}\partial
_{\psi _{j}}\mathcal{K}$, which, with the help of Eq. (\ref{Ka}), are shown
to have the form%
\begin{gather*}
f_{12}=-g\psi _{3}+4\chi _{12}\psi _{1}\psi _{2}, \\
f_{13}=-g\psi _{2}+4\chi _{13}\psi _{1}\psi _{3}, \\
f_{23}=-g\psi _{1}+4\chi _{23}\psi _{2}\psi _{3}, \\
f_{ii}=4\chi _{ii}n_{i}+2\sum_{j=1}^{3}\chi _{ij}n_{j}-2\mu _{i}.
\end{gather*}%
Note that we have defined the partial derivatives in the Hessian matrix to
be taken with respect to $\psi _{i}$ instead of $n_{i}$ so that we can carry
out the stability analysis of both pure and mixed phases in a unified
manner. \

\begin{figure}[tbh]
\includegraphics[width=8cm]{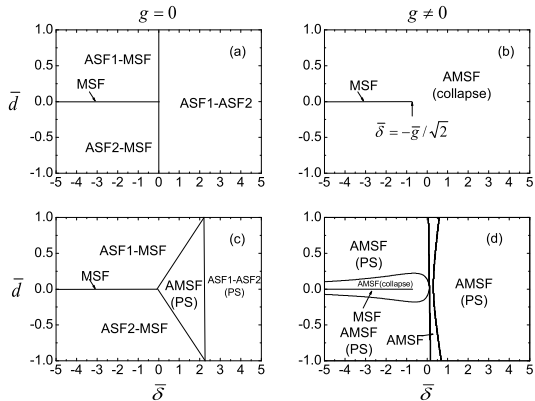}
\caption{{\protect\footnotesize The phase diagrams in the }$\protect\delta $%
{\protect\footnotesize -}$d${\protect\footnotesize \ space where (a) and (b)
are produced with }$\bar{g}=0${\protect\footnotesize \ for (a) and }$\bar{g}%
=1${\protect\footnotesize \ for (b), and (c) and (d) correspond respectively
to (a) and (b) while collisional coefficients are set at }$\bar{\protect\chi}%
_{11}=0.2${\protect\footnotesize , }$\bar{\protect\chi}_{22}=0.3$%
{\protect\footnotesize , }$\bar{\protect\chi}_{12}=2,${\protect\footnotesize %
\ and }$\bar{\protect\chi}_{3i}=0$. {\protect\footnotesize See the text for
units. For the sake of clarity, the labels for ASF1 and ASF2, which are
located at $\bar{d}=1$ and $\bar{d}=-1,$ respectively, are omitted.}}
\label{fig1}
\end{figure}

A state (phase) is thermodynamically stable if it is a local minimum of the
function $\mathcal{K}$ for a given set of chemical potentials. This requires
the Hessian matrix to be positive semidefinite, which, for a symmetric
matrix as in Eq. (\ref{hessian}), is possible if and only if all the
principal minors of Hessian matrix are nonnegative. This leads to a
hierarchy of stability conditions \
\begin{subequations}
\label{stability}
\begin{gather}
f_{ii}\geq 0,  \label{stability 1} \\
f_{ii}f_{jj}-f_{ij}^{2}\geq 0,\text{ }\left( i\neq j\right) ,
\label{stability 2} \\
f_{11}f_{22}f_{33}+2f_{12}f_{13}f_{23}  \notag \\
-f_{11}f_{23}^{2}-f_{22}f_{13}^{2}-f_{33}f_{12}^{2}\geq 0,
\label{stability 3}
\end{gather}%
which reflects the underling philosophy that for a three-component miscible
system to be stable, all the subsets with single, double, and triple
components must also independently satisfy the relevant stability conditions
\cite{ueda06}.

\section{Results and Discussions}

In this section, we construct phase diagrams and discuss phase separations
under a variety of conditions. \ In all the numerical examples, we use two
quantities, $n_{0}$ with the dimension of the density and $\mathcal{E}%
_{0}=\hbar ^{2}n_{0}^{2/3}\left( m_{1}+m_{2}\right) /m_{1}m_{2}$ with the
dimension of the energy, to define a scaling which consists of the following
set of dimensionless variables: $\mathcal{\bar{K}}=\mathcal{K}$/$\mathcal{E}%
_{0}n_{0},$ $\bar{n}_{i}=n_{i}/n_{0},$ $\bar{\psi}_{i}=\psi _{i}/\sqrt{n_{0}}%
,$ $\bar{\mu}_{i}=\mu _{i}/\mathcal{E}_{0},\bar{\delta}=\delta /\mathcal{E}%
_{0}$, $\bar{g}=g\sqrt{n_{0}}/\mathcal{E}_{0}$, and $\bar{\chi}_{ij}=\chi
_{ij}n_{0}/\mathcal{E}_{0}$. \ To gain a familiarity with the parameters, we
consider the interspecies Feshbach resonance around $35$ $G$ in the $^{41}K$-%
$^{87}Rb$ mixture, which is characterized with a magnetic field width $%
\Delta B=5.1$ $G$, a magnetic moment difference $\Delta \mu =0.005\mu _{B}$ (%
$\mu _{B}$ being the Bohr magneton), and a background scattering length $%
a_{bg}=284a_{0}$ ($a_{0}$ being the Bohr radius) \cite{inguscio08,modugno08}%
. \ For a typical density $n_{0}=10^{15}$ $cm^{-3}$, the energy unit is
around $\mathcal{E}_{0}\approx 2.4\times 10^{-29}$ $J$. Thus, $g=\sqrt{2\pi
\hbar ^{2}a_{bg}\Delta B\Delta \mu /m_{12}}$, when converted into the
unitless form, becomes $\bar{g}\approx 0.96$. Additionally, based on $%
a_{11}=99a_{0}$, $a_{22}=60a_{0}$, and $a_{12}=640a_{0}$ \cite{modugno08}
where we have designated $^{87}Rb$ as species 1 and $^{41}K$ as species 2,
we estimate that\ $\bar{\chi}_{11}\approx 0.2$, $\bar{\chi}_{22}\approx 0.3$%
, and $\bar{\chi}_{12}\approx 2.13$. \

In our discussions below, we first map out the phase diagrams in the
chemical potential space. This includes solving Eqs. (\ref{psi123}) to
obtain different phases and to perform stability analysis according to the
criteria (\ref{stability}). These diagrams enable us to identify the regions
where different phases overlap and hence help to answer the question of into
which phases an unstable homogeneous system will separate. However, from the
experimental point of view, more useful are the phase diagrams in the $%
\delta $-$d$ ($\equiv n_{1}-n_{2})$ space, which are directly accessible to
experiments via two control "knobs", the Feshbach detuning $\delta $ and the
two species atomic condensate population difference $d$. \

\subsection{Collisionless Limit}

In order to identify the roles of collisions more clearly in later sections,
we first consider a situations where we artificially turn off all the $s$%
-wave scatterings. \ For pedagogical reasons, here we only present the phase
diagrams in the $\delta $-$d$ space, which are in practice translated from
the phase diagrams in the chemical potential space (see the example in Sec. %
\ref{Sec:collisions} for more details regarding the mapping between the two
types of diagrams.)

\subsubsection{Without Feshbach Coupling $\left( g=0\right) $}

In the collisionless limit and without the Feshbach coupling ($g=0$), our
system simplifies to a three-species ideal gas model, where Eqs. (\ref%
{psi123}) reduces to
\end{subequations}
\begin{equation}
\mu _{i}\psi _{i}=0.  \label{simple equation}
\end{equation}%
Figure \ref{fig1}(a) is the corresponding phase diagram in the $\delta $-$d$
space. This model is simple enough that the features in Fig. \ref{fig1}(a)
are almost self-explanatory. Take as an example the doubly mixed phase
ASF1-MSF with $\psi _{1}\neq 0,$ $\psi _{3}\neq 0,$ and $\psi _{2}=0$. Such
a phase is possible only when $\mu _{1}=\mu _{3}=0$ but $\mu _{2}$ is
arbitrary according to Eqs. (\ref{simple equation}). This means that $d=\psi
_{1}^{2}-\psi _{2}^{2}=\psi _{1}^{2}>0$ and $\delta =\mu _{1}+\mu _{2}-\mu
_{3}=\mu _{2}$\thinspace \thinspace $<0$, where the inequality is due to the
stability criterion (\ref{stability 1}). This is why the top left region of
Fig. \ref{fig1}(a) hosts ASF1-MSF. Other features can be similarly
explained. We note that Fig. \ref{fig1}(a) has similar properties as in Fig.
5 of Ref. \cite{yabu07}. The physics is very clear: as the Feshbach detuning
is tuned from positive to negative or equivalently as the molecular energy
level is lowered from above to below the atomic levels, the system makes a
transition from the atomic side with atomic phases to the molecular side
dominated with phases involving the molecular component.

\

\subsubsection{With finite Feshbach Coupling $\left( g\neq 0\right) $}

Figure \ref{fig1}(b) is the phase diagram in the $\delta $-$d$ space, when
the Feshbach coupling is turned on. Noticeable changes, compared to Fig. \ref%
{fig1}(a) where $g=0$, are the disappearance of the doubly mixed phases into
the AMSF phase, and the corresponding transition boundaries into an
atom-molecule crossover region (see Sec. \ref{Sec:possiblePhase} for the
explanation). In addition, the MSF transition is shifted to $\bar{\delta}=-%
\bar{g}/\sqrt{2}$, owing to the modification of the molecular binding energy
by the Feshbach coupling \cite{radzihovsky04,romans04,duine04}.

We explain below that without collisions, the AMSF phase will collapse. Let
us begin with Eqs. (\ref{psi123}) which, in the absence of collisions,
reduce to $2\mu _{1}\psi _{1}=-g\psi _{2}\psi _{3},$ $2\mu _{2}\psi
_{2}=-g\psi _{1}\psi _{3},$ $2\mu _{3}\psi _{3}=-g\psi _{1}\psi _{2}$. Thus,
within the chemical potential space defined by
\begin{equation}
\mu _{1}<0,\mu _{2}<0\text{, }\mu _{3}<0\text{ }(\text{or }\mu _{1}+\mu
_{2}<\delta ),  \label{chemical condition}
\end{equation}%
there exists a physical AMSF solution ($\psi _{i}>0$) with a density
distribution%
\begin{equation}
n_{1}=\frac{4\mu _{2}\mu _{3}}{g^{2}},\text{ }n_{2}=\frac{4\mu _{1}\mu _{3}}{%
g^{2}},\text{ }n_{3}=\frac{4\mu _{1}\mu _{2}}{g^{2}}.
\label{density distribution}
\end{equation}%
This AMSF phase is unstable because the left-hand side of the stability
condition (\ref{stability 3}) is found to be $32\mu _{1}\mu _{2}\mu _{3}$,
which turns out to be less than zero as a result of the condition (\ref%
{chemical condition}). The only other homogeneous solution besides this AMSF
is the vacuum, which is stable within the space defined by Eq.~(\ref%
{chemical condition}) according to Eq. (\ref{stability 1}). Thus, we
conclude that the only route for this unstable AMSF to take is to collapse.
The system may, however, be stabilized by kinetic energy without going into
a complete collapse, exhibiting the so-called rarified liquid-like behavior
\cite{timmermans99}. The dynamics of phase separation and collapse may lead
to quite complex spatial patterns \cite{ronen}.

\subsection{Collisions Involving Atoms Only}

\label{Sec:collisions}

In general, collisions greatly complicate the construction and understanding
of the phase diagrams, because Eqs. (\ref{psi123}) then become highly
nonlinear and may support several AMSFs. In light of the fact that the
collisional coefficients involving the molecular state are not known in most
experiments, we ignore in this subsection all the collisions involving the
molecular component $\left( \chi _{i3}=0\right) $, and take into
consideration only the atomic collisions. In particular, we focus our
discussion on the case with $\bar{\chi}_{11}=0.2$ and $\bar{\chi}_{22}=0.3$,
and $\bar{\chi}_{12}=2$, where the interspecies collision is far stronger
than the intraspecies collisions. \

\subsubsection{Without Feshbach Coupling $\left( g=0\right) $}

Figure \ref{fig1}(c) is the phase diagram in the $\delta $-$d$ space when $%
g=0$. Figure \ref{fig1}(c) contains, in addition to all the phases occurred
in Fig. \ref{fig1}(a) where collisions are turned off, a unique AMSF given
by
\begin{equation}
n_{1}=\frac{\chi _{12}\mu _{2}-\chi _{22}\mu _{1}}{\chi _{12}^{2}-\chi
_{11}\chi _{22}},n_{2}=\frac{\chi _{12}\mu _{1}-\chi _{11}\mu _{2}}{\chi
_{12}^{2}-\chi _{11}\chi _{22}},
\end{equation}%
$n_{3}=\left( 1-n_{1}-n_{2}\right) /2$, and $\mu _{3}=0$. This AMSF is
surrounded by three lines,
\begin{eqnarray*}
d &=&-\frac{\delta }{\chi _{12}+\chi _{22}}, \\
d &=&\frac{\delta }{\chi _{12}+\chi _{11}}, \\
d &=&\frac{2\delta -\chi _{11}-2\chi _{12}-\chi _{22}}{\chi _{11}-\chi _{22}}%
,
\end{eqnarray*}%
serving as the borders between AMSF and MSF, \ ASF1-MSF, and ASF1-ASF2,
respectively. These equations are obtained from the conditions that $n_{i}=0$
($i=1,2,3$), respectively. Evidently, all the lines are collapsed into a
vertical line at $\delta =0$ and no AMSF is possible in the absence of
collisions as in Fig. \ref{fig1}(a). \ Understandably, ASF1-ASF2 and its
next level of hierarchy, AMSF, are both expected to be unstable against
phase separation due to the strong repulsion between the two atomic
components \cite{timmermans98,ueda06}.

\subsubsection{With finite Feshbach Coupling $\left( g\neq 0\right) $}

In the case of finite Feshbach coupling, we first obtain from Eqs.~(\ref%
{psi123}) that
\begin{equation}
\psi _{3}=-\frac{g}{2\mu _{3}}\psi _{1}\psi _{2},  \label{AMSF psi3}
\end{equation}%
which, when substituted into the other two equations, gives a unique AMSF
solution with the following atom density distribution
\begin{equation}
n_{1}=\frac{\chi _{22}\mu _{1}-\chi _{12}^{\prime }\mu _{2}}{\chi _{11}\chi
_{22}-\chi _{12}^{\prime 2}},\text{ }n_{2}=\frac{\chi _{11}\mu _{2}-\chi
_{12}^{\prime }\mu _{1}}{\chi _{11}\chi _{22}-\chi _{12}^{\prime 2}},
\label{AMSF n1 n2}
\end{equation}%
where $\chi _{12}^{\prime }=\chi _{12}+g^{2}/4\mu _{3}$ plays the role of an
effective interaction between the two atomic species. The extra term $%
g^{2}/4\mu _{3}$ can be traced to the Feshbach coupling; an atom of species
1 appears to interact with an atom of species 2 via the molecular state by
the Feshbach coupling of the strength $g$. This interaction becomes virtual
when $\delta $ is large but can become substantial when $\delta $ is near
resonance. Unlike the AMSF in the case of $g=0$, where $\mu _{3}=0$, $\mu
_{3}$ here must be less than zero ($\mu _{3}<$ $0$). Thus, we always have $%
\chi _{12}^{\prime }<\chi _{12}$. \ It is well known that a double BEC
mixture is stable when $\chi _{12}^{\prime 2}<\chi _{11}\chi _{22}$, and
otherwise, it is unstable and will separate when $\chi _{12}^{\prime }>0$,
and collapse when $\chi _{12}^{\prime }<0$ \cite{timmermans98,ueda06}.

Figure \ref{fig1}(d) shows the phase diagram in the $\delta $-$d$ space when
$g$ is finite. For the reason given previously, we observe that the regions
of doubly mixed phases give away to the AMSF crossover region. With our
choice of $\chi _{ij}$, the system has a natural tendency to phase separate.
However, we see that a narrow strip of stable AMSF emerges from the region
near the Feshbach resonance. This is possible because $\chi _{12}^{\prime }$
can be substantially smaller than $\chi _{12}$ while remaining positive.
This illustrates how one can change the mixture from immiscible to miscible
by varying the Feshbach detuning \cite{wieman08}.

In addition, in the region around the MSF line, because most populations
will accumulate in the molecular component, $\mu _{3}=-g\psi _{1}\psi
_{2}/2\psi _{3}$ will take a small negative value, which in turn makes the
contribution from $g^{2}/4\mu _{3}$ to the effective interparticle
interaction $\chi _{12}^{\prime }$ relatively strong. \ So the system will
collapse in this region but phase separate outside this region where the
repulsive interaction become dominant again.

\subsection{Collisions Involving Both Atoms and Molecules}

The main new feature when collisions involving molecules are also included
is that more than one AMSFs may emerge. To demonstrate this feature, we
consider, for simplicity, a model in which a finite $\chi _{33}$ is
introduced in addition to the atomic collisional terms (while still keeping $%
\chi _{13}=\chi _{23}=0$). By solving Eqs. (\ref{psi123}), we find that the
AMSF phase is governed by equations same as Eqs. (\ref{AMSF psi3}) and (\ref%
{AMSF n1 n2}) except that $\mu _{3}$ is now replaced with $\mu _{3}^{\prime
} $ $=\mu _{3}-\chi _{33}n_{3}$. As can be seen, the chemical potential is
now modulated by a Kerr nonlinear term $-\chi _{33}n_{3}$. It is well known
that the Kerr nonlinear term can lead to multiple solutions that can result
in multi-stability in nonlinear systems. Here, we find (not shown) that $%
n_{3}$ is governed by a fifth-order polynomial equation, and hence may
support more than one physical solutions.

\begin{figure}[tbh]
\includegraphics[width=8cm]{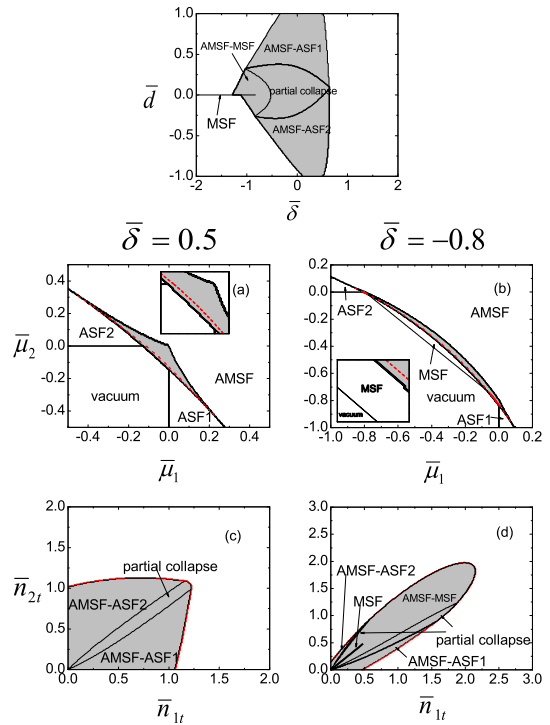}
\caption{{\protect\footnotesize (Color online) The phase diagrams in the }$%
\protect\delta -d${\protect\footnotesize \ space (the upmost figure, with
total density $\bar{n}_{t}=1$), the chemical potential space [(a) and (b)],
and the density space [(c) and (d)] when }$\bar{g}=1${\protect\footnotesize %
, }$\bar{\protect\chi}_{11}=0.2${\protect\footnotesize , }$\bar{\protect\chi}%
_{22}=0.3${\protect\footnotesize , }$\bar{\protect\chi}_{33}=0.245,$%
{\protect\footnotesize \ }$\bar{\protect\chi}_{12}=0.1225$%
{\protect\footnotesize , and }$\bar{\protect\chi}_{13}=\bar{\protect\chi}%
_{23}=0${\protect\footnotesize . Additionally, (a) and (c) are produced with
}$\bar{\protect\delta}=0.5${\protect\footnotesize \ while (b) and (d) are
produced with }$\bar{\protect\delta}=-0.8${\protect\footnotesize . The gray
area in (a) and (b) represents the region where two stable phases can
coexist. The red dashed lines inside the gray area are the first-order phase
transition lines at which the ground state changes from one phase to the
other. See the inset for more details of the region around the first-order
line. Consult the text for units and for a detailed explanation of each
diagram.}}
\label{fig2}
\end{figure}

\subsubsection{One Stable AMSF and One Unstable AMSF}

Let us consider here a case where we choose the molecule-molecule
collisional coefficient to be $\bar{\chi}_{33}=0.245$, and all the atomic
collisional parameters same as in Figs. \ref{fig1}(c) and (d) except that
the interspecies repulsion is now set at a moderate value of $\bar{\chi}%
_{12}=0.1225$. The phase diagram in the $\delta $-$d$ space for this case is
displayed in the upmost figure of Fig. \ref{fig2}, which is all occupied by
the AMSF state except for the line at $d=0$ marked with MSF. \ Because of
the much reduced $\chi _{12}$, the system has a natural tendency to be in a
miscible state and the AMSF will not phase separate (the white area) except
in a small parameter region (the grey area) close to the Feshbach resonance.
Unlike in Fig. \ref{fig1} where we continue to label the phase separated
region as AMSF, in Fig. \ref{fig2}, we will not show the unstable AMSF,
instead, we divide it into regions and label them according to how the
homogeneous (but unstable) AMSF is going to phase separate. \ For example,
AMSF-ASF1 means that the homogeneous AMSF within the grey region of that
label will phase separate into one AMSF and one ASF1.

The origin of this rich set of phase separations can be best understood from
the chemical potential and density space. Consider, for example, $\bar{\delta%
}=0.5$. In Fig. \ref{fig2}(a), we construct the phase diagram in the
chemical potential space. \ For a fixed set of chemical potentials, the
coupled equations (\ref{psi123}) are then found to support two physical
AMSFs in some parameter regimes: one is stable and the other is unstable.
Only the stable AMSF is shown here. \ As Fig. \ref{fig2}(a) illustrates, the
stable AMSF coexists and shares the first-order transition line,
respectively, with ASF1, ASF2, and vacuum. The first-order transition line
in Fig. \ref{fig2}(a) is transformed into the grey region in Fig. \ref{fig2}%
(c) when the phase diagram is mapped from the chemical potential to the
density space \cite{sheehy07,marchetti08}. \ \ For our purpose, we find it
convenient to define the densities $n_{1t}=n_{1}+n_{3}$ and $%
n_{2t}=n_{2}+n_{3}$ so that $n_{1t}+n_{2t}$ and $n_{1t}-n_{2t}\left(
=d\right) $ represent the total density and the population imbalance,
respectively. For a system operating within the grey region in Fig. \ref%
{fig2}(c), it will phase separate into AMSF and ASF1, or AMSF and ASF2, and
or AMSF and vacuum. The separated phases in each set have different
densities, and phase separation is a natural course for a system to take in
order to maintain the average densities specified within the respective
domain of the phase separation. The phase separation into AMSF-vacuum
signals collapse.

As $\delta $ is reduced, for example, to $\bar{\delta}=-0.8$, a new phase
separation, AMSF-MSF, appears in the upmost figure, which is again supported
by Figs. \ref{fig2}(b) and (d) in the chemical potential and density space.
Compared to $\bar{\delta}=0.5$ where the AMSF is unstable irrespective of
the value of $d$, the AMSF at $\bar{\delta}=-0.8$ becomes a stable one when
the population imbalance between the two atomic species is sufficiently
large.

\begin{figure}[tbh]
\includegraphics[width=8cm]{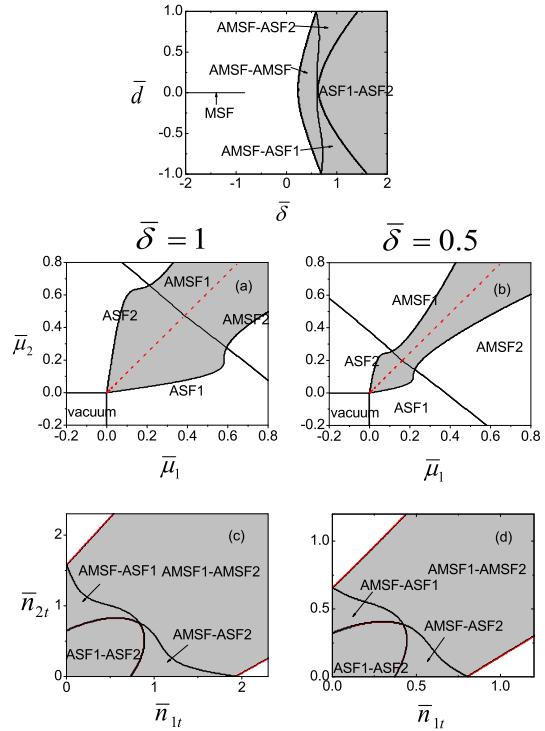}
\caption{{\protect\footnotesize (Color online) The phase diagrams similarly
defined as in Fig. 2. \ The parameters are the same as those in Fig. 2
except that }$\bar{\protect\chi}_{12}=2${\protect\footnotesize . \
Additionally, (a) and (c) are produced with }$\bar{\protect\delta}=1$%
{\protect\footnotesize \ while (b) and (d) are produced with }$\bar{\protect%
\delta}=0.5${\protect\footnotesize . \ The red dashed lines have the same
meanings as that in Fig. 2. Consult the text for units.}}
\label{fig3}
\end{figure}

\subsubsection{Two Stable AMSFs and One Unstable AMSF}

\label{sec:twoAMSFs}

The nonlinearity of Eqs. (\ref{psi123}) may provide the opportunity for an
unstable homogeneous AMSF to phase separate into two different AMSFs.
Consider the model that shares the same parameters as in Fig. \ref{fig2}
except that $\chi _{12}$ is set back to a value $\left( \bar{\chi}%
_{12}=2\right) $ same as in Fig. \ref{fig1}. It is found that Eqs. (\ref%
{psi123}) support three physical AMSF solutions, two of which satisfy the
stability conditions (\ref{stability}) and are hence stable. The phase
diagrams corresponding to this model are shown in Fig. \ref{fig3}. A
coexistence region between two stable AMSFs having different population
distributions is identified from the chemical potential space both for $\bar{%
\delta}=1$ [Fig. \ref{fig3}(a)] and $\bar{\delta}=0.5$ [Fig. \ref{fig3}(b)].
However, in experiments where the total population is fixed, AMSF1-AMSF2 is
accessible only to sufficiently small Feshbach detunings, for example, $\bar{%
\delta}=0.5$ (see the phases crossed by the (imagined) line $\bar{n}_{1t}+%
\bar{n}_{2t}=1$ in both Fig. \ref{fig3}(c) and Fig. \ref{fig3}(d)]. \ We
emphasize that so far, we have not seen reports of phase separation into two
AMSFs both in theory and in experiments.\textbf{\ }In particular, we mention
that the homonuclear systems are known to only support phase separations
involving one AMSF \cite{radzihovsky04}. For this reason, we suspect that
this may be a phenomenon quite unique to the heteronuclear model.

This case also provides an example for us to see the effect of the
molecule-molecule interaction on the stability of the AMSF. \ Compared to
Fig. \ref{fig1}(d) where $\chi _{33}=0$, the upmost diagram of Fig. \ref%
{fig3} shows that the presence of $\chi _{33}$ stabilized the AMSF on the
molecular side of the Feshbach resonance.

\section{Summary}

In conclusion, we have studied the zero-temperature mean-field phase
diagrams of two-species atomic BEC mixtures with the interspecies Feshbach
resonance under a variety of conditions, with a special attention being
given to the subject of phase separation. We have found that on one hand,
the Feshbach coupling alone cannot create a stable AMSF; to create stable
AMSF, collisions must be present, and on the other hand, the Feshbach
coupling may stabilize a system that has a natural tendency to phase
separate under collisions. We have shown that the molecule-molecule
collision can stabilize the AMSF on the molecular side of the Feshbach
resonance. We have identified the interspecies Feshbach interaction to be
the cause for the absence of the doubly mixed phases, as well as the reason
for the presence of both pure atom and pure molecule phases in our model. \
This provides the opportunity, as we have indeed verified numerically, for
observing the phase separation not only between AMSF and MSF but also those
between AMSF and pure ASFs. We have also studied the collapse or partial
collapse of the AMSF as special phase separations where one of the separated
components is the vacuum. Under certain conditions, we have even found that
our system is able to phase separate into two distinct AMSFs, which, we
speculate, is a property unique to the heteronuclear model.

\acknowledgments

This work is supported by the National Natural Science Foundation of China
under Grant No. 10588402 and No. 10474055, the National Basic Research
Program of China (973 Program) under Grant No. 2006CB921104, the Science and
Technology Commission of Shanghai Municipality under Grant No. 06JC14026 and
No. 05PJ14038, the Program of Shanghai Subject Chief Scientist under Grant
No. 08XD14017, the Program for Changjiang Scholars and Innovative Research
Team in University, Shanghai Leading Academic Discipline Project under Grant
No. B480, the Research Fund for the Doctoral Program of Higher Education
under Grant No. 20040003101 (W.Z.), by China Postdoctoral Science Foundation
under Grant No. 44021570 (L.Z.) and by the US National Science Foundation
(H.P., H.Y.L.), the US Army Research Office (H.Y.L.), and the Robert A.
Welch Foundation (Grant No. C-1669), and the W. M. Keck Foundation (H.P.).

$^{\dagger }$wpzhang@phy.ecnu.edu.cn

$^{\ast }$ling@rowan.edu

$^{\ast \ast }$hpu@rice.edu

\end{document}